\documentclass{ws-mpla}
\pdfoutput=1
\usepackage[T1]{fontenc}
\usepackage{bm,amsmath,amssymb,color,bbm,mathrsfs} 
\usepackage[super]{cite}
\usepackage{xcolor}
\usepackage[verbose,hypertexnames=false]{hyperref}
\hypersetup{colorlinks=false,allbordercolors=blue,pdfborderstyle={/S/U/W 1}}

\DeclareMathOperator\diag{diag}
\DeclareMathOperator\Tr{Tr}
\DeclareMathOperator\Pf{Pf}
\DeclareMathOperator\rme{\mathrm{e}}

\renewcommand{\bar}[1]{\overline{#1}}

\newcommand{\bep}{\begin{pmatrix}} 
\newcommand{\eep}{\end{pmatrix}}
\newcommand{\USp}{\text{USp}}
\newcommand{\SU}{\text{SU}}

\newcommand{\U}{\text{U}}
\newcommand{\1}{\mathbbm{1}}

\newcommand{\CC}{\mathbb{C}}

\renewcommand{\epsilon}{\varepsilon}
\newcommand{\rmd}{\mathrm{d}}

\newcommand{\Z}{{\mathcal{Z}}}
\newcommand{\M}{\hat{M}}
\newcommand{\m}{\hat{m}}

\def\ba#1\ea{\begin{align}#1\end{align}}
\def\mkakko#1{\left( #1 \right)}
\def\ckakko#1{\left\{ #1 \right\}}
\def\kkakko#1{\left[ #1 \right]}
\def\wt#1{\widetilde{#1}}

\begin{document}

\markboth{T.~Kanazawa}{Relativistic Cooper pairing in random matrix theory}

%%%%%%%%%%%%%%%%%%%%% Publisher's Area please ignore %%%%%%%%%%%%%%
\catchline{}{}{}{}{}
%%%%%%%%%%%%%%%%%%%%%%%%%%%%%%%%%%%%%%%%%%%%%%%%%%%%%%%%%%%%%%%%%%%

\title{Relativistic Cooper pairing in the microscopic limit of chiral random matrix theory}

\author{Takuya Kanazawa}
\address{Department of Business Administration, Kobe Gakuin University, 1-1-3 Minatojima, Chuo-ku, Kobe 650-8586, Japan
\\
tkanazawa@ba.kobegakuin.ac.jp}

\allowdisplaybreaks
\maketitle

\begin{history}
\received{(Day Month Year)}
\revised{(Day Month Year)}
\accepted{(Day Month Year)}
\published{(Day Month Year)}
\end{history}

\begin{abstract}
Random matrix theory (RMT) provides a powerful framework for analyzing universal features of strongly coupled physical systems. In quantum chromodynamics (QCD), cold quark matter at asymptotically high density is expected to exhibit color superconductivity (CSC), the analogue of superconductivity in condensed-matter systems. Although CSC phases have been studied within RMT primarily in the macroscopic large-$N$ limit, where $N$ denotes the matrix size, it has remained unclear whether an RMT exists that realizes CSC in the microscopic large-$N$ limit. Here we answer this question in the affirmative by introducing a novel non-Hermitian chiral random matrix model. For three quark flavors, we show that the model exhibits spontaneous breaking of color $\mathrm{SU}(3)$ and flavor $\mathrm{SU}(3)$ symmetries down to the diagonal $\mathrm{SU}(3)$ subgroup, thereby reproducing color-flavor locking in dense QCD. For two flavors, we find that color $\mathrm{SU}(3)$ is spontaneously broken to $\mathrm{SU}(2)$ while the chiral symmetry $\mathrm{SU}(2)_{\mathrm{L}}\times\mathrm{SU}(2)_{\mathrm{R}}$ remains unbroken, consistent with the two-flavor color-superconducting phase.
\end{abstract}

\keywords{Chiral symmetry; color superconductivity; random matrices}

\ccode{PACS Nos.: 02.10.Yn, 12.39.Fe, 21.65.Qr}

\section{Introduction}

Random matrix theory (RMT) has found broad applications in mathematics, physics, statistics, biology, information theory, and engineering \cite{Mehtabook,AGZbook,Akemann:2011csh,CDbook}. In physics, RMT has long been used to describe universal energy-level fluctuations of quantum-chaotic Hamiltonians \cite{Guhr:1997ve,Haakebook}. In nuclear and elementary particle physics, RMT with the same global symmetries as QCD has been employed to elucidate various aspects of dynamical symmetry breaking in low-energy hadron physics \cite{Halasz:1998qr,Verbaarschot:2000dy,Vanderheyden:2011iq}. It is well established that spontaneous chiral symmetry breaking in QCD with massless quarks is intimately tied to the accumulation of near-zero eigenvalues of the Dirac operator \cite{Banks:1979yr}. Spectral fluctuations of the Dirac eigenvalues on the scale $\sim 1/(V_4\Sigma)$, where $V_4$ is the spacetime volume and $\Sigma$ denotes the chiral condensate $|\langle\bar\psi\psi\rangle|$, are universal and can be described \emph{exactly} by chirally symmetric RMT. This limit of QCD, in which the partition function is dominated by the zero modes of the Nambu--Goldstone fields, is known as the $\epsilon$-regime \cite{Gasser:1987ah,Leutwyler:1992yt,Osborn:1998qb}. It is mathematically equivalent to the \emph{microscopic} large-$N$ limit of chiral RMT \cite{Shuryak:1992pi,Verbaarschot:1997bf,Verbaarschot:2000dy,Verbaarschot:2005rj}. In contrast, there have also been attempts to take the \emph{macroscopic} large-$N$ limit of chiral RMT to probe symmetry breaking and phase transitions in QCD. The goal of the latter is to construct a phenomenological Ginzburg--Landau-type thermodynamic potential for QCD in terms of macroscopic order parameters, without focusing on the universal fluctuations of Dirac eigenvalues at the scale $1/(V_4\Sigma)$. By contrast, chiral RMT in the microscopic limit magnifies this particular region of the spectrum and is not designed to capture the global phase diagram of QCD. In this sense, the two approaches are complementary. Eigenvalue distribution functions obtained analytically from chiral RMT in the microscopic limit have been used to determine $\Sigma$ in lattice QCD simulations \cite{PhysRevD.76.054503}.

If one wishes to investigate nuclear and quark matter in the interior of compact stars, one must introduce a chemical potential $\mu$; the Dirac operator then ceases to be anti-Hermitian, and its eigenvalues spread over the complex plane. A natural question is whether the powerful framework of RMT can be applied to QCD at nonzero chemical potential. In the $\varepsilon$-regime, i.e., $\mu\to0$ and $V_4\to\infty$ with $V_4F_\pi^2\mu^2\sim 1$, where $F_\pi$ is the pion decay constant, substantial progress has been made \cite{Akemann:2007rf}. RMTs that explicitly incorporate $\mu$ were proposed \cite{Stephanov:1996ki,Osborn:2004rf}, and the microscopic spectral functions were worked out in full detail \cite{Splittorff:2003cu,Osborn:2004rf,Akemann:2004dr}.

On the other hand, constructing a microscopic RMT applicable to the dense regime $\mu\ne0$ is highly nontrivial, because simply taking $\mu$ large in the models of Refs.~\citen{Stephanov:1996ki,Osborn:2004rf} renders them trivial, with neither spectral fluctuations nor symmetry breaking. Progress was made in Ref.~\citen{Kanazawa:2009en}, where an RMT describing two-color QCD at high density $\mu\gg\Lambda_{\rm QCD}$ was identified. The exact microscopic spectral functions of this model were derived analytically \cite{Akemann:2009fc,Akemann:2010mt,Akemann:2010tv}. Later, the construction of Ref.~\citen{Kanazawa:2009en} was generalized to adjoint QCD at high density and to QCD with large isospin chemical potential \cite{Kanazawabook}. A Banks--Casher-type relation linking the density of complex Dirac eigenvalues at the origin to the BCS gap $\Delta$ was also established for these QCD-like theories \cite{Kanazawa:2012zr} and tested in lattice simulations \cite{Brandt:2019hel}. Furthermore, RMT for the \emph{singular-value} spectrum of the Dirac operator was studied in Ref.~\citen{Kanazawa:2011tt}. A common feature of these QCD-like theories is that the high-density condensate is a color singlet and therefore does not induce color superconductivity (CSC).

By contrast, in three-color QCD, diquarks are not color singlets, and the color symmetry is spontaneously broken via the Higgs mechanism at large quark chemical potential \cite{Bailin:1983bm}. Many CSC phases have been proposed on the QCD phase diagram \cite{Rajagopal:2000wf,Alford:2007xm,Fukushima:2010bq,Casalbuoni:2018haw}. In particular, in QCD with three degenerate flavors, the ground state at asymptotically high density and low temperature is believed to be the color-flavor-locked (CFL) phase \cite{Alford:1998mk}, characterized by the diquark condensates
\ba
	\langle\psi^a_{f}\sigma_2\psi^b_{g}\rangle \propto
	\kappa_1\delta_{af}\delta_{bg} + \kappa_2\delta_{ag}\delta_{bf}
	\label{eq:qqcfl}
\ea
where $a,b$ are color indices and $f,g$ are flavor indices. This condensate breaks $\SU(3)_{\rm color}\times\SU(3)_{\rm L}\times\SU(3)_{\rm R}$ down to the diagonal $\SU(3)_{\rm V}$ subgroup. All quarks and gluons are gapped, and the low-energy dynamics is dominated by $8+1+1$ Nambu--Goldstone modes%
\footnote{$8$ comes from $\SU(3)_{\rm A}$ and 1 from $\U(1)_{\rm B}$. The last 1 comes from $\U(1)_{\rm A}$; the axial anomaly is suppressed at high density \cite{Schafer:2002ty}.}. The $\epsilon$-regime is well defined for the CFL phase, and one can derive infinitely many spectral sum rules for the complex Dirac eigenvalues on the scale $\sim 1/\sqrt{V_4\Delta^2}$ \cite{Yamamoto:2009ey}. Apart from this result, essentially nothing is known about statistical properties of the Dirac operator at high quark density.%
\footnote{There is, however, recent progress \cite{Kanazawa:2020ktn,Kanazawa:2020fpo} that incorporates CSC into RMT in the microscopic limit.} Since lattice simulations are plagued by the notorious sign problem, this regime is currently inaccessible to direct numerical studies.

In this paper, we construct a new RMT that describes dynamical locking of color and flavor symmetries%
\footnote{In RMT there is no local gauge invariance and all symmetries are global. Hence the distinction between color and flavor is a formal one.}
induced by diquark condensates of the form \eqref{eq:qqcfl}. We rigorously take the microscopic large-$N$ limit and determine the quark-mass dependence of the partition function. We show analytically that the effective action for the soft modes begins at second order in the quark masses, implying that the chiral condensate vanishes. We further show that the quark-mass dependence is governed by fluctuations of a soft mode taking values in $\U(3)$, which can be interpreted as a color-singlet four-quark field $\bar{\psi}_{\rm L}\bar\psi_{\rm L}\psi_{\rm R} \psi_{\rm R} +\text{h.c.}$ arising from the coupling between a left-handed diquark and a right-handed diquark. This is consistent with the physical picture of pions in the CFL phase \cite{Fukushima:2004bj}.

A salient feature of our RMT is the absence of an explicit parameter corresponding to the chemical potential, in stark contrast to previous works \cite{Stephanov:1996ki,Halasz:1998qr,Vanderheyden:2000ti,Osborn:2004rf,Sano:2011xs}. The random matrix for the left-handed sector is statistically independent of that for the right-handed sector, rendering the Dirac matrix \emph{maximally non-Hermitian}. Such a decoupling of left- and right-handed quarks in the chiral limit indeed occurs in high-density QCD \cite{Alford:2007xm}, and it is therefore natural that our RMT reflects this feature.%
\footnote{We note that RMT for dense two-color QCD also had this feature \cite{Kanazawa:2009en}.}

This paper is organized as follows. In Section~\ref{sc:rmtdef}, we define the new matrix model and summarize its basic properties. In Section~\ref{sc:largeN}, we rewrite the model for $N_f=3$ using Hubbard--Stratonovich fields and take the large-$N$ limit. We demonstrate color-flavor locking and derive the corresponding sigma-model representation. In Section~\ref{sc:twosc}, we analyze the model for $N_f=2$, derive the color-symmetry breaking pattern $\SU(3)\to\SU(2)$, and discuss its implications for the 2SC phase of QCD. We conclude in Section~\ref{sc:conc}.

\section{\label{sc:rmtdef}Matrix model}

The RMT we propose is defined by the partition function
\ba
	\hspace{-3mm}
	Z & = \int \prod_{A=0}^{8}\rmd V^A \int \prod_{A=0}^{8}\rmd W^A \int \rmd X \int \rmd Y 
	\prod_{f=1}^{N_f}\det(\mathscr{D}+m_f \1_{12N})
	\notag\\
	& \quad \times
	\exp\big[-2N \Tr(V^{A\dagger}V^A)-2N \Tr(W^{A\dagger}W^A)
	-N\Tr (X^\dagger X)-N\Tr (Y^\dagger Y)\big]
\ea
where the ``Dirac operator'' $\mathscr{D}$ is a $12N\times 12N$ matrix defined as
\ba
	\mathscr{D} = \begin{pmatrix}
	0 & D_{\rm R}
	\\
	D_{\rm L} & 0
	\end{pmatrix}
\ea
with
\ba
	D_{\rm L} & \equiv 
	(CV^A+V^{A*}C) \otimes \lambda^A + i (CX + X^*C)\otimes \1_3\,,
	\\
	D_{\rm R} & \equiv 
	(CW^A+W^{A*}C) \otimes \lambda^A + i (CY + Y^*C)\otimes \1_3\,.
\ea
Here, $V^A,W^A(A=0,1,\cdots,8), X$ and $Y$ are $2N\times 2N$ complex matrices with independently and identically distributed matrix elements, $C=i\sigma_2 \otimes \1_N$ is a $2N\times 2N$ skew-symmetric martix, $\lambda^{\mathfrak{a}}(\mathfrak{a}=1,\cdots,8)$ are $3\times 3$ Gell-Mann matrices normalized as $\Tr(\lambda^{\mathfrak{a}}\lambda^{\mathfrak{b}})=2\delta_{\mathfrak{a}\mathfrak{b}}$, and $\lambda^0=\sqrt{\frac{2}{3}}\1_3$. $\rmd V^A, \rmd W^A, \rmd X$ and $\rmd Y$ are flat Cartesian measures. Since $D_{\rm L}$ and $D_{\rm R}$ are statistically independent, $\mathscr{D}$ is maximally non-Hermitian. In the chiral limit, the model is clearly invariant under $\U(N_f)_{\rm L}\times\U(N_f)_{\rm R}$. In addition, there are \emph{two} ``color'' $\SU(3)_{\rm C}$ symmetries that act on the left and right sectors separately. They are locked to the diagonal $\SU(3)$ subgroup by quark masses. There is also a symplectic symmetry
\begin{gather}
	V^A \to g V^A h^\dagger, \quad X \to g X h^\dagger
	\label{eq:666}
	\\
	W^A \to h^*W^A g^T, \quad Y \to h^*Yg^T
	\label{eq:777}
\end{gather}
where $g,h\in\USp(2N)$, namely
\ba
	g^T C g = h^T C h = C,~~g^\dagger g=h^\dagger h = \1_{2N}\,.
\ea
The skew-symmetric matrix $C$ is a matrix-model counterpart of the Pauli matrix $\sigma_2$ in \eqref{eq:qqcfl}, which contracts spinor indices to make the diquark a Lorentz scalar. $C$ does not show up in the standard chiral RMT \cite{Shuryak:1992pi,Verbaarschot:2000dy} because the chiral condensate $\bar\psi\psi$ requires no skew-symmetrization of spinor indices.

\section{\label{sc:largeN}Three flavors}

Let us cast the model for $N_f=3$ to the form that is more amenable to the large-$N$ analysis. First, we introduce quarks $\psi_{f\alpha}^{a}$ and anti-quarks $\bar\psi_{f\alpha}^{a}$. Here and in the following, we label colors of quarks by $a,b,c,d,e\in\{1,2,3\}$ and flavors by $f,g,h,i,j\in\{1,2,3\}$. The Greek indices $\alpha,\beta,\gamma,\delta$ run from 1 to $2N$. The quark mass matrix is defined as $M=\diag(m_1,m_2,m_3)$. Then the determinant can be expressed as a Grassmannian integral
\ba
	Z & = \int \rmd \bar\psi_{\rm R} \rmd \bar\psi_{\rm L} \rmd \psi_{\rm R} \rmd \psi_{\rm L} \int \prod_{A=0}^{8}\rmd V^A \int \prod_{A=0}^{8}\rmd W^A \int \rmd X \int \rmd Y 
	\notag
	\\
	& \quad \times \exp\big[-2N \Tr(V^{A\dagger}V^A)-2N \Tr(W^{A\dagger}W^A)
	-N\Tr (X^\dagger X)-N\Tr (Y^\dagger Y)
	\notag
	\\
	& \quad + \left.
	\scalebox{0.75}{$\displaystyle 
	\begin{pmatrix}\bar\psi_{\rm R} \\ \bar\psi_{\rm L}
	\end{pmatrix}^a_{f\alpha}\hspace{-5pt}\begin{pmatrix}
	m_f\delta_{\alpha\beta}\delta_{ab} & (CW^A + W^{A*}C)_{\alpha\beta}\lambda^A_{ab} + i(CY+Y^*C)_{\alpha\beta}\delta_{ab} 
	\\ 
	(CV^A+V^{A*}C)_{\alpha\beta}\lambda^A_{ab} + i (CX + X^*C)_{\alpha\beta}\delta_{ab} & m^*_f \delta_{\alpha\beta}\delta_{ab}
	\end{pmatrix}\begin{pmatrix}
	\psi_{\rm L} \\ \psi_{\rm R}
	\end{pmatrix}^b_{f\beta}
	$} \right]
	\\
	& = \int \rmd \bar\psi_{\rm R} \rmd \bar\psi_{\rm L} \rmd \psi_{\rm R} \rmd \psi_{\rm L} \int \prod_{A=0}^{8}\rmd V^A \int \prod_{A=0}^{8}\rmd W^A \int \rmd X \int \rmd Y 
	\notag
	\\
	& \quad \times 
	\exp\Big[
		\bar\psi^a_{{\rm R}\alpha}M\psi^a_{{\rm L}\alpha} + \bar\psi^a_{{\rm L}\alpha}M^\dagger \psi^a_{{\rm R}\alpha}
	\notag
	\\
	& \qquad 
	- N(2V^{A*}_{\alpha\beta}V^A_{\alpha\beta}	
	+ X^{*}_{\alpha\beta}X_{\alpha\beta}
	+ 2W^{A*}_{\alpha\beta}W^A_{\alpha\beta}
	+ Y^{*}_{\alpha\beta}Y_{\alpha\beta}	)
	\notag
	\\
	& \qquad 
	+\bar\psi^a_{{\rm L}f\alpha}(C_{\alpha\gamma}V^A_{\gamma\beta}+V^{A*}_{\alpha\gamma}C_{\gamma\beta})\lambda^A_{ab}\psi^b_{{\rm L}f\beta}
	+i\bar\psi^a_{{\rm L}f\alpha}(C_{\alpha\gamma}X_{\gamma\beta}+X^*_{\alpha\gamma}C_{\gamma\beta})\psi^a_{{\rm L}f\beta}
	\notag
	\\
	& \qquad 
	+\bar\psi^a_{{\rm R}f\alpha}(C_{\alpha\gamma}W^A_{\gamma\beta}+W^{A*}_{\alpha\gamma}C_{\gamma\beta})\lambda^A_{ab}\psi^b_{{\rm R}f\beta}
	+i\bar\psi^a_{{\rm R}f\alpha}(C_{\alpha\gamma}Y_{\gamma\beta}+Y^*_{\alpha\gamma}C_{\gamma\beta})\psi^a_{{\rm R}f\beta}
	\Big]\,.
	\hspace{-2mm}
\ea
To keep invariance under \eqref{eq:666} and \eqref{eq:777}, the Grassmann fields must transform as
\ba
	\psi_{\rm L}\to h \psi_{\rm L}, \quad 
	\bar\psi_{\rm L}\to \bar\psi_{\rm L}g^T, \quad 
	\psi_{\rm R}\to g^* \psi_{\rm R}, \quad 
	\bar\psi_{\rm R} \to \bar\psi_{\rm R} h^\dagger
	\label{eq:834jj}
\ea
for $g,h\in\USp(2N)$.

It is tedious but straightforward to integrate out the Gaussian random matrices $V^A, W^A, X$ and $Y$. The result reads
\ba
	Z & \propto \int \rmd \bar\psi_{\rm R} \rmd \bar\psi_{\rm L} \rmd \psi_{\rm R} \rmd \psi_{\rm L} 
	\exp\bigg[\bar\psi^a_{{\rm R} \alpha}M\psi^a_{{\rm L}\alpha} + \bar\psi^a_{{\rm L}\alpha}M^\dagger \psi^a_{{\rm R}\alpha}
	\notag
	\\
	& \qquad + 
	\frac{1}{2N}(\bar\psi^a_{{\rm L}f\gamma}C_{\gamma\alpha}\lambda^A_{ab}\psi^b_{{\rm L}f\beta})
	(\bar\psi^c_{{\rm L}g\alpha}C_{\beta\delta}\lambda^A_{cd}\psi^d_{{\rm L}g\delta})
	\notag
	\\
	& \qquad 
	-\frac{1}{N}(\bar\psi^a_{{\rm L}f\gamma}C_{\gamma\alpha}\psi^a_{{\rm L}f\beta})
	(\bar\psi^b_{{\rm L}g\alpha}C_{\beta\delta}\psi^b_{{\rm L}g\delta}) 
	+ ({\rm L}\leftrightarrow {\rm R})
	\bigg]\,.
\ea

Next we use the relation $\lambda^A_{ab}\lambda^A_{cd}=2\delta_{ad}\delta_{bc}$ to obtain
\ba
	\hspace{-3mm}
	Z & \propto \int \rmd \bar\psi_{\rm R} \rmd \bar\psi_{\rm L} \rmd \psi_{\rm R} \rmd \psi_{\rm L} 
	\exp\bigg[\bar\psi^a_{{\rm R}\alpha}M\psi^a_{{\rm L}\alpha} + \bar\psi^a_{{\rm L}\alpha}M^\dagger \psi^a_{{\rm R}\alpha}
	\notag
	\\
	& \quad + 
	\frac{1}{N}(\bar\psi^a_{{\rm L}f\gamma}C_{\gamma\alpha}\psi^b_{{\rm L}f\beta})
	(\bar\psi^b_{{\rm L}g\alpha}C_{\beta\delta}\psi^a_{{\rm L}g\delta})
	-\frac{1}{N}(\bar\psi^a_{{\rm L}f\gamma}C_{\gamma\alpha}\psi^a_{{\rm L}f\beta})
	(\bar\psi^b_{{\rm L}g\alpha}C_{\beta\delta}\psi^b_{{\rm L}g\delta})
	\notag
	\\
	& \quad 
	+ ({\rm L}\leftrightarrow {\rm R})
	\bigg]
	\\
	& = \int \rmd \bar\psi_{\rm R} \rmd \bar\psi_{\rm L} \rmd \psi_{\rm R} \rmd \psi_{\rm L}
	\exp\bigg[\bar\psi^a_{{\rm R}\alpha}M\psi^a_{{\rm L}\alpha} + \bar\psi^a_{{\rm L}\alpha}M^\dagger \psi^a_{{\rm R}\alpha}
	\notag
	\\
	& \quad + 
	\frac{1}{N}(\delta_{ac}\delta_{bd} - \delta_{ad}\delta_{bc})\delta_{fh}\delta_{gi}
	(\bar\psi^a_{{\rm L}f\gamma}C_{\gamma\alpha}\bar\psi^b_{{\rm L}g\alpha})
	(\psi^c_{{\rm L}h\beta}C_{\beta\delta}\psi^d_{{\rm L}i\delta}) + ({\rm L}\leftrightarrow {\rm R})
	\bigg]
	\\
	& = \int \rmd \bar\psi_{\rm R} \rmd \bar\psi_{\rm L} \rmd \psi_{\rm R} \rmd \psi_{\rm L} 
	\exp\bigg[\bar\psi^a_{{\rm R}\alpha}M\psi^a_{{\rm L}\alpha} + \bar\psi^a_{{\rm L}\alpha}M^\dagger \psi^a_{{\rm R}\alpha}
	\notag
	\\
	& \quad + 
	\frac{1}{2N}\epsilon_{abe}\epsilon_{cde}(\delta_{fh}\delta_{gi}-\delta_{fi}\delta_{gh})
	(\bar\psi^a_{{\rm L}f\gamma}C_{\gamma\alpha}\bar\psi^b_{{\rm L}g\alpha})
	(\psi^c_{{\rm L}h\beta}C_{\beta\delta}\psi^d_{{\rm L}i\delta}) 
	+ ({\rm L} \leftrightarrow {\rm R})
	\bigg]
	\label{eq:genflav}
	\\
	& = \int \rmd \bar\psi_{\rm R} \rmd \bar\psi_{\rm L} \rmd \psi_{\rm R} \rmd \psi_{\rm L} 
	\exp\bigg[\bar\psi^a_{{\rm R}\alpha}M\psi^a_{{\rm L}\alpha} + \bar\psi^a_{{\rm L}\alpha}M^\dagger \psi^a_{{\rm R}\alpha}
	\notag
	\\
	& \quad + 
	\frac{1}{2N}\epsilon_{abe}\epsilon_{cde}\epsilon_{fgj}\epsilon_{hij}
	(\bar\psi^a_{{\rm L}f\gamma}C_{\gamma\alpha}\bar\psi^b_{{\rm L}g\alpha})(\psi^c_{{\rm L}h\beta}C_{\beta\delta}\psi^d_{{\rm L}i\delta}) 
	+ ({\rm L} \leftrightarrow {\rm R})
	\bigg]\,.
	\label{eq:oierwsnfsdk}
\ea
To bilinearize the quartic interaction, we insert the constant factor
\begin{align}
	& \int \rmd \Delta_{\rm L}
	\exp\bigg[
		-2N\ckakko{(\Delta_{\rm L})^*_{ej}-\frac{1}{2N}\epsilon_{cde}\epsilon_{hij}\psi^c_{{\rm L}h\beta}C_{\beta\delta}\psi^d_{{\rm L}i\delta}}
	\notag
	\\
	& \qquad \times 
	\ckakko{(\Delta_{\rm L})_{ej}-\frac{1}{2N}\epsilon_{abe}\epsilon_{fgj}\bar\psi^a_{{\rm L}f\gamma}C_{\gamma\alpha}\bar\psi^b_{{\rm L}g\alpha}}
	\bigg]
	\notag
	\\
	& \times \int \rmd \Delta_{\rm R}
	\exp\bigg[
		-2N\ckakko{(\Delta_{\rm R})^*_{ej}-\frac{1}{2N}\epsilon_{cde}\epsilon_{hij}\psi^c_{{\rm R}h\beta}C_{\beta\delta}\psi^d_{{\rm R}i\delta}}
	\notag
	\\
	& \qquad \times 
	\ckakko{(\Delta_{\rm R})_{ej}-\frac{1}{2N}\epsilon_{abe}\epsilon_{fgj}\bar\psi^a_{{\rm R}f\gamma}C_{\gamma\alpha}\bar\psi^b_{{\rm R}g\alpha}}
	\bigg]
\end{align}
where $\Delta_{\rm L,R}$ are $3\times 3$ complex matrices. They transform as triplet under both color $\SU(3)$ and flavor $\SU(3)$, but singlet under the symplectic symmetry \eqref{eq:834jj}. 
Then
\ba
	Z & \propto \int \rmd \Delta_{\rm L} \int \rmd \Delta_{\rm R} \int \rmd \bar\psi_{\rm R} \rmd \bar\psi_{\rm L} \rmd \psi_{\rm R} \rmd \psi_{\rm L}
	\exp\Big[\bar\psi^a_{{\rm R}\alpha}M\psi^a_{{\rm L}\alpha} + \bar\psi^a_{{\rm L}\alpha}M^\dagger \psi^a_{{\rm R}\alpha}
	\notag
	\\
	& \qquad -2N \Tr (\Delta_{\rm L}^\dagger \Delta_{\rm L})-2N \Tr (\Delta_{\rm R}^\dagger \Delta_{\rm R}) 
	\notag
	\\
	& \qquad 
	+ \epsilon_{cde}\epsilon_{hij}\psi^c_{{\rm L}h\beta}C_{\beta\delta}\psi^d_{{\rm L}i\delta}\Delta_{{\rm L}ej}
	+ \epsilon_{abe}\epsilon_{fgj}\bar\psi^a_{{\rm L}f\gamma}C_{\gamma\alpha}\bar\psi^b_{{\rm L}g\alpha} \Delta^*_{{\rm L}ej}
	\notag
	\\
	& \qquad 
	+ \epsilon_{cde}\epsilon_{hij}\psi^c_{{\rm R}h\beta}C_{\beta\delta}\psi^d_{{\rm R}i\delta}\Delta_{{\rm R}ej}
	+ \epsilon_{abe}\epsilon_{fgj}\bar\psi^a_{{\rm R}f\gamma}C_{\gamma\alpha}\bar\psi^b_{{\rm R}g\alpha}\Delta^*_{{\rm R}ej}
	\Big]
	\\
	& = \int \rmd \Delta_{\rm L} \int \rmd \Delta_{\rm R} \int \rmd \bar\psi_{\rm R} \rmd \bar\psi_{\rm L} \rmd \psi_{\rm R} \rmd \psi_{\rm L} 
	\exp\big[
	-2N \Tr (\Delta_{\rm L}^\dagger \Delta_{\rm L})-2N \Tr (\Delta_{\rm R}^\dagger \Delta_{\rm R})
	\notag
	\\
	& \quad + \left.
	\scalebox{0.8}{$\displaystyle 
	\begin{pmatrix}\bar\psi_{\rm L}\\\bar\psi_{\rm R}\\\psi_{\rm L}\\\psi_{\rm R}\end{pmatrix}^a_{f\alpha}
	\hspace{-6pt}
	\begin{pmatrix}
	\epsilon_{abc}\epsilon_{fgh}\Delta^*_{{\rm L}ch}C_{\alpha\beta}&0&0&(M^\dagger)_{fg}\frac{\delta_{ab}\delta_{\alpha\beta}}{2}
	\\
	0& \epsilon_{abc}\epsilon_{fgh}\Delta^*_{{\rm R}ch}C_{\alpha\beta} &M_{fg}\frac{\delta_{ab}\delta_{\alpha\beta}}{2}&0
	\\
	0&-(M^T)_{fg}\frac{\delta_{ab}\delta_{\alpha\beta}}{2}& \epsilon_{abc}\epsilon_{fgh}\Delta_{{\rm L}ch}C_{\alpha\beta} &0
	\\
	-M^*_{fg}\frac{\delta_{ab}\delta_{\alpha\beta}}{2}
	&0&0& \epsilon_{abc}\epsilon_{fgh}\Delta_{{\rm R}ch}C_{\alpha\beta}\end{pmatrix}
	\begin{pmatrix}\bar\psi_{\rm L}\\\bar\psi_{\rm R}\\\psi_{\rm L}\\\psi_{\rm R}\end{pmatrix}^b_{g\beta}
	$} \right] .
\ea
Now one can integrate out quarks and obtain a product of Pfaffians:
\ba
	Z & \propto \int \rmd \Delta_{\rm L} \int \rmd \Delta_{\rm R}  
	\exp\kkakko{-2N \Tr (\Delta_{\rm L}^\dagger \Delta_{\rm L})-2N \Tr (\Delta_{\rm R}^\dagger \Delta_{\rm R})}
	\notag
	\\
	& \qquad \times 
	\Pf\begin{pmatrix}\epsilon_{abc}\epsilon_{fgh}\Delta^*_{{\rm L}ch}C_{\alpha\beta} & (M^\dagger)_{fg}\frac{\delta_{ab}\delta_{\alpha\beta}}{2} 
	\\ 
	-M^*_{fg}\frac{\delta_{ab}\delta_{\alpha\beta}}{2} & \epsilon_{abc}\epsilon_{fgh}\Delta_{{\rm R}ch}C_{\alpha\beta}
	\end{pmatrix}
	\notag
	\\
	& \qquad \times 
	\Pf\begin{pmatrix}
	\epsilon_{abc}\epsilon_{fgh}\Delta^*_{{\rm R}ch}C_{\alpha\beta} &M_{fg}\frac{\delta_{ab}\delta_{\alpha\beta}}{2}
	\\
	-(M^T)_{fg}\frac{\delta_{ab}\delta_{\alpha\beta}}{2} & \epsilon_{abc}\epsilon_{fgh}\Delta_{{\rm L}ch}C_{\alpha\beta}
	\end{pmatrix}
	\\
	& = \int \rmd \Delta_{\rm L} \int \rmd \Delta_{\rm R}  
	\exp\kkakko{-2N \Tr (\Delta_{\rm L}^\dagger \Delta_{\rm L})-2N \Tr (\Delta_{\rm R}^\dagger \Delta_{\rm R})}
	\notag
	\\
	& \qquad \times {\det}^{N}\begin{pmatrix}
	\epsilon_{abc}\epsilon_{fgh}\Delta^*_{{\rm L}ch} &(M^\dagger)_{fg}\frac{\delta_{ab}}{2} 
	\\
	-M^*_{fg}\frac{\delta_{ab}}{2} &-\epsilon_{abc}\epsilon_{fgh}\Delta_{{\rm R}ch}
	\end{pmatrix}
	\notag
	\\
	& \qquad \times 
	{\det}^{N}\begin{pmatrix}
	\epsilon_{abc}\epsilon_{fgh}\Delta^*_{{\rm R}ch}&M_{fg}\frac{\delta_{ab}}{2}
	\\
	-(M^T)_{fg}\frac{\delta_{ab}}{2}
	&-\epsilon_{abc}\epsilon_{fgh}\Delta_{{\rm L}ch}
	\end{pmatrix}.
\ea
This is an exact rewriting of the original partition function and so far no approximation has been made. We are now ready to take the microscopic large-$N$ limit in which $M\to 0$ and $N\to \infty$ such that $NM^2\sim 1$. The first step is to determine the saddle point in the chiral limit. Setting $M=0$, we find
\ba
	Z & \propto \int \rmd \Delta_{\rm L} \int \rmd \Delta_{\rm R}  
	\exp \big[
	-2N \Tr (\Delta_{\rm L}^\dagger \Delta_{\rm L})-2N \Tr (\Delta_{\rm R}^\dagger \Delta_{\rm R})
	\big]
	\notag
	\\
	& \qquad \times {\det}^N(\epsilon_{abc}\epsilon_{fgh}\Delta^*_{{\rm L}ch})
	{\det}^N(\epsilon_{abc}\epsilon_{fgh}\Delta_{{\rm L}ch})
	\notag
	\\
	& \qquad \times {\det}^N(\epsilon_{abc}\epsilon_{fgh}\Delta^*_{{\rm R}ch})
	{\det}^N(\epsilon_{abc}\epsilon_{fgh}\Delta_{{\rm R}ch}) \,.
\ea
Note that $\epsilon_{abc}\epsilon_{fgh}\Delta_{ch}$ is regarded here as a $9\times 9$ symmetric matrix with a left index $(a,f)$ and a right index $(b,g)$. It holds that%
\footnote{We used SymPy, a Python library for symbolic mathematics \cite{10.7717/peerj-cs.103}.}
\ba
	\det (\epsilon_{abc}\epsilon_{fgh}\Delta_{ch}) = -2(\det \Delta)^3 \,.
\ea
Hence
\ba
	Z & \propto \int \rmd \Delta_{\rm L} \int \rmd \Delta_{\rm R}  
	\exp\big[
	-2N \Tr (\Delta_{\rm L}^\dagger \Delta_{\rm L})-2N \Tr (\Delta_{\rm R}^\dagger \Delta_{\rm R})
	\big]
	\notag
	\\
	& \qquad \times {\det}^{3N}(\Delta_{\rm L}^\dagger \Delta_{\rm L}){\det}^{3N}(\Delta_{\rm R}^\dagger \Delta_{\rm R}) \,.
\ea
A singular value decomposition yields $\Delta_{\rm L}=u\Lambda_{\rm L}v$ with $\Lambda_{\rm L}$ a diagonal matrix with non-negative entries and $u,v\in\U(3)$, and likewise for $\Delta_{\rm R}$. It then follows that the saddle point at large $N$ is located at
\ba
	\Lambda_{\rm L}=\Lambda_{\rm R}=\sqrt{\frac{3}{2}}\,\1_3\,.
\ea
This means that the $\SU(3)$ color symmetry and the $\SU(3)$ flavor symmetry are spontaneously broken to the diagonal $\SU(3)$ subgroup, i.e., \emph{color-flavor locking}. 

The soft mode around the saddle point can be parametrized in terms of unitary matrices $U$ and $V$ as
\ba
	\Delta_{\rm L} = \sqrt{\frac{3}{2}}\,U,\quad 
	\Delta_{\rm R} = \sqrt{\frac{3}{2}}\,V \,.
\ea
(This $V$ should not be confused with the random matrix $V^A$ appearing in the Dirac operator $\mathscr{D}$.)  
Now we reinstate the quark masses and find
\ba
	Z & \approx 
	\int_{\U(3)}\!\!\!\! \rmd U \int_{\U(3)}\!\!\!\! \rmd V ~ 
	{\det}^{N}\begin{pmatrix}
	\epsilon_{abc}\epsilon_{fgh}U^*_{ch} &(M^\dagger)_{fg}\frac{\delta_{ab}}{\sqrt{6}} 
	\\
	-M^*_{fg}\frac{\delta_{ab}}{\sqrt{6}} &-\epsilon_{abc}\epsilon_{fgh}V_{ch}
	\end{pmatrix}
	\notag
	\\
	& \qquad \times 
	{\det}^{N}\begin{pmatrix}
	\epsilon_{abc}\epsilon_{fgh}V^*_{ch}&M_{fg}\frac{\delta_{ab}}{\sqrt{6}}
	\\
	-(M^T)_{fg} \frac{\delta_{ab}}{\sqrt{6}} & -\epsilon_{abc}\epsilon_{fgh}U_{ch}
	\end{pmatrix}\,.
\ea
The evaluation of these determinants needs some preparation. 
Let us define 
\ba
	\mathcal{U}_{af,bg}&\equiv \epsilon_{abc}\epsilon_{fgh}U_{ch}, 
	\\
	\mathcal{V}_{af,bg}&\equiv \epsilon_{abc}\epsilon_{fgh}V_{ch},
	\\
	\wt{U}_{af,bg} &\equiv U_{ab}\delta_{fg},
	\\
	\wt{V}_{af,bg} &\equiv V_{ab}\delta_{fg},
	\\
	\mathcal{M}_{af,bg} & \equiv M_{fg}\delta_{ab},
	\\
	\Z_{af,bg}& \equiv \epsilon_{abc}\epsilon_{fgc},
	\\
	\rme^{i\phi_U} & \equiv \det U,
	\\
	\rme^{i\phi_V} & \equiv \det V. 
\ea 
Then it holds that
\ba
	\wt{U}^\dagger \mathcal{U}^* \wt{U}^* & = \rme^{-i\phi_U} \Z\,,
	\label{eq:UUU}
	\\
	\wt{V}^T\mathcal{V}\wt{V} & = \rme^{i\phi_V} \Z\,,
	\label{eq:VVV}
	\\
	\wt{V}^\dagger \mathcal{M} \wt{U} & = (V^\dagger U) \otimes M\,.
	\label{eq:VMU}
\ea
The proof is straightforward. For instance, 
\ba
	(\wt{V}^T\mathcal{V}\wt{V})_{af,bg} & = 
	(\wt{V}^T)_{af,a'f'}\mathcal{V}_{a'f',b'g'}\wt{V}_{b'g',bg}
	\\
	& = V_{a'a}\delta_{ff'}\epsilon_{a'b'c}\epsilon_{f'g'h}V_{ch}V_{b'b}\delta_{g'g}
	\\
	& = \epsilon_{fgh}(\epsilon_{a'b'c}V_{a'a}V_{b'b}V_{ch})
	\\
	& = \epsilon_{fgh}\epsilon_{abh}\det V
	\\
	& = (\det V)\Z_{af,bg}\,.
\ea
Using \eqref{eq:UUU}, \eqref{eq:VVV} and \eqref{eq:VMU} we find
\ba
	& \det 
	\begin{pmatrix}
	\mathcal{U}^* & \mathcal{M}^\dagger /\sqrt{6} 
	\\
	-\mathcal{M}^*/\sqrt{6} & -\mathcal{V}
	\end{pmatrix}
	\det 
	\begin{pmatrix}
	\mathcal{V}^* & \mathcal{M}/\sqrt{6} 
	\\
	-\mathcal{M}^T/\sqrt{6} & -\mathcal{U}
	\end{pmatrix}
	\notag
	\\
	=\; & 
	\det \kkakko{
	\begin{pmatrix}\wt{U}^\dagger &0\\0&\wt{V}^T\end{pmatrix}
	\begin{pmatrix}
	\mathcal{U}^* & \mathcal{M}^\dagger /\sqrt{6} 
	\\
	-\mathcal{M}^*/\sqrt{6} & -\mathcal{V}
	\end{pmatrix}
	\begin{pmatrix}\wt{U}^*&0\\0&\wt{V}\end{pmatrix}}
	\notag
	\\
	& \quad \times  
	\det\kkakko{
	\begin{pmatrix}\wt{V}^\dagger &0\\0&\wt{U}^T\end{pmatrix}
	\begin{pmatrix}
	\mathcal{V}^* & \mathcal{M}/\sqrt{6} 
	\\
	-\mathcal{M}^T/\sqrt{6} & -\mathcal{U}
	\end{pmatrix}
	\begin{pmatrix}\wt{V}^*&0\\0&\wt{U}\end{pmatrix}}
	\\
	=\; & \det \begin{pmatrix}
		\wt{U}^\dagger \mathcal{U}^* \wt{U}^* & 
		\wt{U}^\dagger \mathcal{M}^\dagger \wt{V}/\sqrt{6}
		\\
		-\wt{V}^T \mathcal{M}^* \wt{U}^*/\sqrt{6} & 
		- \wt{V}^T \mathcal{V} \wt{V}
	\end{pmatrix}
	\det\begin{pmatrix}
		\wt{V}^\dagger \mathcal{V}^*\wt{V}^* & 
		\wt{V}^\dagger \mathcal{M} \wt{U}/\sqrt{6}
		\\
		-\wt{U}^T\mathcal{M}^T\wt{V}^*/\sqrt{6}&
		-\wt{U}^T\mathcal{U}\wt{U}
	\end{pmatrix}
	\\
	=\; & \det\begin{pmatrix}
	\rme^{-i\phi_U} \Z & (U^\dagger V)\otimes M^\dagger / \sqrt{6}\\
	- (V^T U^*)\otimes M^* / \sqrt{6}&-\rme^{i\phi_V}\Z\end{pmatrix}
	\notag
	\\
	& \quad \times 
	\det\begin{pmatrix}
	\rme^{-i\phi_V}\Z & (V^\dagger U)\otimes M / \sqrt{6}
	\\
	-(U^TV^*)\otimes M^T / \sqrt{6} &-\rme^{i\phi_U}\Z\end{pmatrix}
	\\
	=\; & \det(-\rme^{-i\phi_U+i\phi_V}\Z^2
	+\Z[(V^T U^*)\otimes M^*]\Z^{-1}[(U^\dagger V)\otimes M^\dagger]/6)
	\notag
	\\
	& \quad \times \det(-\rme^{i\phi_U-i\phi_V}\Z^2
	+\Z[(U^TV^*)\otimes M^T]\Z^{-1}[(V^\dagger U)\otimes M]/6)
	\\
	\propto\; & \det\Big(\1_9 - \rme^{i(\phi_U-\phi_V)}
	\Z^{-1}[(V^T U^*)\otimes M^*]\Z^{-1}[(U^\dagger V)\otimes M^\dagger]/6\Big)
	\notag
	\\
	& \quad \times \det\Big(\1_9 - \rme^{-i(\phi_U-\phi_V)} 
	\Z^{-1}[(U^TV^*)\otimes M^T]\Z^{-1}[(V^\dagger U)\otimes M]/6 \Big)\,.
\ea
Now we take the scaling limit $N\to \infty$ with $\M \equiv \sqrt{N}M$ fixed. The result is
\ba
	Z \sim 
	& \int_{\U(3)}\!\!\!\! \rmd \hat{U} ~ \exp\bigg(-\frac{1}{6}\Tr\ckakko{\Z^{-1}(\hat{U}^T\otimes \M^T)\Z^{-1}(\hat{U}\otimes \M)}
	+\text{c.c.}\bigg)
	\label{eq:zzzumum}
\ea
where we have defined $\hat{U}\equiv\rme^{-i(\phi_U-\phi_V)/2}V^\dagger U$. Using SymPy~\cite{10.7717/peerj-cs.103} we find
\ba
	& \Tr\ckakko{\Z^{-1}(\hat{U}^T\otimes \M^T)\Z^{-1}(\hat{U}\otimes \M)} 
	= \frac{5}{4}[\Tr(\M \hat{U})]^2 - \Tr[(\M\hat{U})^2]\,.
\ea
Thus
\ba
	Z & \sim \int_{\U(3)}\!\!\!\! \rmd \hat{U} ~ \exp\bigg(-\frac{1}{6}\ckakko{\frac{5}{4}[\Tr(\M \hat{U})]^2 - \Tr[(\M\hat{U})^2]}
	+\text{c.c.}\bigg).
	\label{eq:ZRMTm}
\ea
This is the sigma-model representation of our new RMT. It is noteworthy that the linear term $\Tr(\M \hat{U})$ is absent, indicating that \emph{the chiral condensate in the massless limit is strictly zero in this model,} and symmetry breaking is instead triggered by diquark condensates. As $U$ ($V$) represents fluctuations of the left-handed (right-handed) diquarks, respectively, the variable $\hat{U}\propto V^\dagger U$ can be interpreted as a color-singet four-quark state, $\bar{\psi}_{\rm R}\bar{\psi}_{\rm R}\psi_{\rm L}\psi_{\rm L} + \text{h.c.}$ 

It is of interest to compare our RMT partition function \eqref{eq:ZRMTm} with the mass term of the chiral effective theory of the CFL phase in dense QCD \cite{Son:1999cm,Son:2000tu,Schafer:2001za},
\ba
	\mathcal{L} = -\frac{3\Delta^2}{4\pi^2}\ckakko{[\Tr(MU)]^2 - \Tr[(MU)^2]} + \text{c.c.}
	\label{eq:cflmassterm}
\ea
The numerical factors in \eqref{eq:ZRMTm} and \eqref{eq:cflmassterm} are slightly different%
\footnote{The CFL mass term is exactly reprodued if we replace $\Z^{-1}$ by $\Z$ in \eqref{eq:zzzumum}.}. Where does this minor discrepancy come from? In the CFL phase of QCD, the diquark condensate of the form \eqref{eq:qqcfl} with $\kappa_1\approx -\kappa_2$ develops. In contrast, our RMT hosts a diquark condensate
\ba
	\langle \psi^a_f \psi^b_g \rangle \propto \frac{1}{2}\delta_{af}\delta_{bg}-\delta_{ag}\delta_{bf}\,.
\ea
This can be verified with a quick calculation. For a squared quark mass term, we have
\ba
	(\bar\psi_{\rm R} M \psi_{\rm L})^2 & \sim \bar\psi^a_{{\rm R}f}\bar\psi^b_{{\rm R}f'}\psi^a_{{\rm L}g}\psi^b_{{\rm L}g'}M_{fg}M_{f'g'}
	\\
	& \sim \mkakko{\frac{1}{2}\delta_{af}\delta_{bf'}-\delta_{af'}\delta_{bf}}
	\mkakko{\frac{1}{2}\delta_{ag}\delta_{bg'}-\delta_{ag'}\delta_{bg}} M_{fg}M_{f'g'}
	\\
	& = \mkakko{\frac{5}{4}\delta_{fg}\delta_{f'g'}-\delta_{fg'}\delta_{f'g}}M_{fg}M_{f'g'}
	\\
	& = \frac{5}{4}(\Tr M)^2 - \Tr (M^2)\,.
\ea
Thus the mass dependence of RMT \eqref{eq:ZRMTm} is reproduced. The bottom line is that \emph{the requirement that color and flavor be locked alone does not uniquely fix the form of the chiral Lagrangian.} Our RMT obeys exactly the same pattern of symmetry breaking as the CFL phase of QCD, and the difference is a purely quantitative one. It is an open problem to modify or generalize our RMT so as to exactly match the CFL chiral Lagrangian \eqref{eq:cflmassterm}, which is left for future work.

\section{\label{sc:twosc}Two flavors}

We now discuss the case of two flavors. We shall begin with \eqref{eq:genflav}, which is valid for an arbitrary number of flavors:
\ba
	\hspace{-6mm}
	Z & \propto \int \rmd \bar\psi_{\rm R} \rmd \bar\psi_{\rm L} \rmd \psi_{\rm R} \rmd \psi_{\rm L} 
	\exp\bigg[\bar\psi^a_{{\rm R}\alpha}M\psi^a_{{\rm L}\alpha} + \bar\psi^a_{{\rm L}\alpha}M^\dagger \psi^a_{{\rm R}\alpha}
	\notag
	\\
	& \qquad + 
	\frac{1}{2N}\epsilon_{abe}\epsilon_{cde}(\delta_{fh}\delta_{gi}-\delta_{fi}\delta_{gh})
	(\bar\psi^a_{{\rm L}f\gamma}C_{\gamma\alpha}\bar\psi^b_{{\rm L}g\alpha})
	(\psi^c_{{\rm L}h\beta}C_{\beta\delta}\psi^d_{{\rm L}i\delta}) 
	\notag
	\\
	& \qquad 
	+ ({\rm L} \leftrightarrow {\rm R})
	\bigg].
\ea
For two flavors there is an identity
\ba
	\epsilon_{fg}\epsilon_{hi}=\delta_{fh}\delta_{gi}-\delta_{fi}\delta_{gh}\,,
\ea
thus 
\ba
	Z & \propto \int \rmd \bar\psi_{\rm R} \rmd \bar\psi_{\rm L} \rmd \psi_{\rm R} \rmd \psi_{\rm L} 
	\exp\bigg[\bar\psi^a_{{\rm R}\alpha}M\psi^a_{{\rm L}\alpha} + \bar\psi^a_{{\rm L}\alpha}M^\dagger \psi^a_{{\rm R}\alpha}
	\notag
	\\
	& \qquad + 
	\frac{1}{2N}\epsilon_{abe}\epsilon_{cde}\epsilon_{fg}\epsilon_{hi}
	(\bar\psi^a_{{\rm L}f\gamma}C_{\gamma\alpha}\bar\psi^b_{{\rm L}g\alpha})
	(\psi^c_{{\rm L}h\beta}C_{\beta\delta}\psi^d_{{\rm L}i\delta}) 
	+ ({\rm L}\leftrightarrow {\rm R})
	\bigg].
\ea
To bilinearize the quartic interaction, we insert the constant factor
\begin{align}
	& \int_{\CC^3} \rmd \Omega_{\rm L} \exp\bigg[
		-2N\mkakko{\Omega_{{\rm L}e}^*-\frac{1}{2N}\epsilon_{cde}\epsilon_{hi}	\psi^c_{{\rm L}h\beta}C_{\beta\delta}\psi^d_{{\rm L}i\delta}}
	\notag
	\\
	& \qquad 
	\times \mkakko{\Omega_{{\rm L}e}-\frac{1}{2N}\epsilon_{abe}\epsilon_{fg}\bar\psi^a_{{\rm L}f\gamma}C_{\gamma\alpha}\bar\psi^b_{{\rm L}g\alpha}}
	\bigg]
	\notag
	\\
	& \times\int_{\CC^3} \rmd \Omega_{\rm R} \exp\bigg[
		-2N\mkakko{\Omega_{{\rm R}e}^*-\frac{1}{2N}\epsilon_{cde}\epsilon_{hi}	\psi^c_{{\rm R}h\beta}C_{\beta\delta}\psi^d_{{\rm R}i\delta}}
	\notag
	\\
	& \qquad 
	\times \mkakko{\Omega_{{\rm R}e}-\frac{1}{2N}\epsilon_{abe}\epsilon_{fg}\bar\psi^a_{{\rm R}f\gamma}C_{\gamma\alpha}\bar\psi^b_{{\rm R}g\alpha}}
	\bigg]
\end{align}
where $\Omega_{\rm L,R}$ are complex three-component vectors that transform as triplet under color $\SU(3)$. 
They are also charged under $\U(1)_{\rm B}$ and $\U(1)_{\rm A}$. Then
\ba
	Z & \propto \int_{\CC^3} \rmd \Omega_{\rm L} \int_{\CC^3} \rmd \Omega_{\rm R} 
	\int \rmd \bar\psi_{\rm R} \rmd \bar\psi_{\rm L} \rmd \psi_{\rm R} \rmd \psi_{\rm L} 
	\exp\bigg[
		-2N(|\Omega_{\rm L}|^2+|\Omega_{\rm R}|^2)  
		\notag
		\\
		& \quad 
		+\bar\psi^a_{{\rm R}\alpha}M\psi^a_{{\rm L}\alpha}
		+ \bar\psi^a_{{\rm L}\alpha}M^\dagger \psi^a_{{\rm R}\alpha}
	\notag
	\\
	& \quad 
	+ \Omega_{{\rm L}c}\epsilon_{abc}\epsilon_{fg}\psi^a_{{\rm L}f\alpha}C_{\alpha\beta}\psi^b_{{\rm L}g\beta}
	+ \Omega^*_{{\rm L}c}\epsilon_{abc}\epsilon_{fg}\bar\psi^a_{{\rm L}f\alpha}C_{\alpha\beta}\bar\psi^b_{{\rm L}g\beta}
	+ ({\rm L} \leftrightarrow {\rm R})
	\bigg]
	\\
	& = \int_{\CC^3} \rmd \Omega_{\rm L} \int_{\CC^3} \rmd \Omega_{\rm R} 
	\int \rmd \bar\psi_{\rm R} \rmd \bar\psi_{\rm L} \rmd \psi_{\rm R} \rmd \psi_{\rm L} 
	\exp \big[ -2N(|\Omega_{\rm L}|^2+|\Omega_{\rm R}|^2)
	\notag
	\\
	& \quad + \left.
	\scalebox{0.8}{$\displaystyle 
	\begin{pmatrix}\bar\psi_{\rm L}\\\bar\psi_{\rm R}\\\psi_{\rm L}\\\psi_{\rm R}\end{pmatrix}^{a}_{f\alpha}
	\begin{pmatrix}
		\Omega^*_{{\rm L}c}\epsilon_{abc}\epsilon_{fg}C_{\alpha\beta}&0&0&
		(M^\dagger)_{fg}\frac{\delta_{ab}\delta_{\alpha\beta}}{2}
		\\
		0&\Omega^*_{{\rm R}c}\epsilon_{abc}\epsilon_{fg}C_{\alpha\beta}&
		M_{fg}\frac{\delta_{ab}\delta_{\alpha\beta}}{2}
		&0
		\\
		0&-(M^T)_{fg}\frac{\delta_{ab}\delta_{\alpha\beta}}{2}
		&	\Omega_{{\rm L}c}\epsilon_{abc}\epsilon_{fg}C_{\alpha\beta}&0
		\\
		-(M^*)_{fg}\frac{\delta_{ab}\delta_{\alpha\beta}}{2}
		&0&0&\Omega_{{\rm R}c}\epsilon_{abc}\epsilon_{fg}C_{\alpha\beta}	
	\end{pmatrix}
	\begin{pmatrix}\bar\psi_{\rm L}\\\bar\psi_{\rm R}\\
	\psi_{\rm L}\\\psi_{\rm R}
	\end{pmatrix}^b_{g\beta}
	$}\right].
\ea
The quarks can be readily integrated out and yield a product of Pfaffians:
\ba
	Z & \propto \int_{\CC^3} \rmd \Omega_{\rm L} \int_{\CC^3} \rmd \Omega_{\rm R} 
	~\rme^{-2N(|\Omega_{\rm L}|^2+|\Omega_{\rm R}|^2)}
	\Pf
	\begin{pmatrix}
		\Omega^*_{{\rm R}c}\epsilon_{abc}\epsilon_{fg}C_{\alpha\beta}&M_{fg}\frac{\delta_{ab}\delta_{\alpha\beta}}{2}
		\\
		-(M^T)_{fg}\frac{\delta_{ab}\delta_{\alpha\beta}}{2}&	\Omega_{{\rm L}c}\epsilon_{abc}\epsilon_{fg}C_{\alpha\beta}
	\end{pmatrix}
	\notag
	\\
	& \qquad \times \Pf
	\begin{pmatrix}
		\Omega^*_{{\rm L}c}\epsilon_{abc}\epsilon_{fg}C_{\alpha\beta} & (M^\dagger)_{fg}\frac{\delta_{ab}\delta_{\alpha\beta}}{2}
		\\
		-(M^*)_{fg}\frac{\delta_{ab}\delta_{\alpha\beta}}{2} & \Omega_{{\rm R}c}\epsilon_{abc}\epsilon_{fg}C_{\alpha\beta}
	\end{pmatrix}
	\\
	& = \int_{\CC^3} \rmd \Omega_{\rm L} \int_{\CC^3} \rmd \Omega_{\rm R} 
	~\rme^{-2N(|\Omega_{\rm L}|^2+|\Omega_{\rm R}|^2)}
	{\det}^N 
	\begin{pmatrix}
		\Omega^*_{{\rm R}c}\epsilon_{abc}\epsilon_{fg} 
		& M_{fg}\frac{\delta_{ab}}{2}
		\\
		-(M^T)_{fg}\frac{\delta_{ab}}{2} 
		& - \Omega_{{\rm L}c}\epsilon_{abc}\epsilon_{fg}
	\end{pmatrix}
	\notag
	\\
	& \qquad \times {\det}^N
	\begin{pmatrix}
		\Omega^*_{{\rm L}c}\epsilon_{abc}\epsilon_{fg} & (M^\dagger)_{fg}\frac{\delta_{ab}}{2}
		\\
		-(M^*)_{fg}\frac{\delta_{ab}}{2} & -\Omega_{{\rm R}c}\epsilon_{abc}\epsilon_{fg}
	\end{pmatrix} 
	\\
	& = \int_{\CC^3} \rmd \Omega_{\rm L} \int_{\CC^3} \rmd \Omega_{\rm R} 
	~\rme^{-2N(|\Omega_{\rm L}|^2+|\Omega_{\rm R}|^2)}
	\notag
	\\
	& \qquad \times {\det}^{2N} 
	\begin{pmatrix}
		\Omega^*_{{\rm R}c}\epsilon_{abc} & m_u \frac{\delta_{ab}}{2}
		\\
		-m_d \frac{\delta_{ab}}{2} & \Omega_{{\rm L}c}\epsilon_{abc}
	\end{pmatrix}
	{\det}^{2N}
	\begin{pmatrix}
		\Omega^*_{{\rm L}c}\epsilon_{abc} & m^*_u\frac{\delta_{ab}}{2}
		\\
		-m_d^* \frac{\delta_{ab}}{2}& \Omega_{{\rm R}c}\epsilon_{abc}
	\end{pmatrix}
	\\
	& = \int_{\CC^3} \rmd \Omega_{\rm L} \int_{\CC^3} \rmd \Omega_{\rm R} 
	~\rme^{-2N(|\Omega_{\rm L}|^2+|\Omega_{\rm R}|^2)}
	\notag
	\\
	& \qquad \times 
	\ckakko{\frac{m_um_d}{4}\mkakko{\Omega_{\rm R}^\dagger\Omega_{\rm L} - \frac{m_um_d}{4}}^2}^{2N}
	\ckakko{\frac{m_u^*m_d^*}{4}\mkakko{\Omega_{\rm L}^\dagger\Omega_{\rm R} - \frac{m_u^*m_d^*}{4}}^2}^{2N}
	\\
	& \propto |m_um_d|^{4N}
	\int_{\CC^3} \rmd \Omega_{\rm L} \int_{\CC^3} \rmd \Omega_{\rm R} 
	~\rme^{-2N(|\Omega_{\rm L}|^2+|\Omega_{\rm R}|^2)}
	\left|\Omega_{\rm R}^\dagger \Omega_{\rm L} - \frac{m_um_d}{4}\right|^{8N}.
	\label{eq:62}
\ea
This is an exact rewriting of the original theory and no approximation has been made yet. 
The fact that the partition function vanishes with high powers of $m$ in the chiral limit indicates that there are macroscopically many gapless quarks in this system, consistent with the physical picture of the 2SC phase \cite{Bailin:1983bm,Alford:1997zt,Rapp:1997zu}. 

Now we shall take the large-$N$ limit. In the chiral limit with masses factored out, we have
\ba
	\frac{Z}{|m_um_d|^{4N}} & \sim 
	\int_{\CC^3} \rmd \Omega_{\rm L} \int_{\CC^3} \rmd \Omega_{\rm R} 
	\rme^{-2N(|\Omega_{\rm L}|^2+|\Omega_{\rm R}|^2)}
	\big|\Omega_{\rm R}^\dagger\Omega_{\rm L}\big|^{8N}.
\ea
Let us define 
\begin{gather}
	\Omega_{\rm L} = \xi v_{\rm L}, \quad \Omega_{\rm R} = \eta v_{\rm R}, 
	\\
	\xi\geq0, ~~\eta\geq0, ~~v_{\rm L}^\dagger v_{\rm L}=v_{\rm R}^\dagger v_{\rm R}=1.
\end{gather}
Then
\ba
	\frac{Z}{|m_um_d|^{4N}} & \sim \int_0^\infty \rmd \xi
	\int_0^\infty \rmd \eta \int \rmd v_{\rm L} \int \rmd v_{\rm R}~
	\rme^{-2N(\xi^2+\eta^2)}(\xi\eta)^{8N}
	|v_{\rm R}^\dagger v_{\rm L}|^{8N}.
\ea
Extremization of the integrand yields 
\ba
	\xi = \eta = \sqrt{2}, ~~ v_{\rm L} = \rme^{i\varphi}v_{\rm R}
	\label{eq:67}
\ea
where $\rme^{i\varphi}$ is an arbitrary $\U(1)$ phase. 
With a suitable $\U(3)$ rotation, $v_{\rm L}$ can be brought to 
$(1,0,0)^T$, hence the saddle point is given by
\ba
	\Omega_{\rm L} = (\sqrt{2},0,0)^T
\ea
and likewise for $\Omega_{\rm R}$. Obviously this breaks the color symmetry spontaneously as
\ba
	\SU(3)_{\rm C} \to \SU(2)_{\rm C} \,,
\ea
but leaves the chiral symmetry $\SU(2)_{\rm L}\times\SU(2)_{\rm R}$ intact. 
It is quite satisfactory to see that this breaking pattern is identical to the 2SC phase of two-flavor QCD at high density \cite{Bailin:1983bm,Alford:1997zt,Rapp:1997zu}. 

We proceed to the derivation of the action for the soft mode $\varphi$, which is the Nambu-Goldstone mode associated with the $\U(1)_{\rm A}$ symmetry. Plugging \eqref{eq:67} into \eqref{eq:62} yields
\ba
	Z & \sim |m_um_d|^{4N} \int_0^{2\pi}\!\!\!\rmd \varphi 
	\left|2\rme^{i\varphi} - \frac{m_um_d}{4}\right|^{8N}.
\ea
For $N\gg1$ with $\m_{u,d}\equiv\sqrt{N}m_{u,d}$ fixed, one gets
\ba
	Z &\sim |\m_u \m_d|^{4N} \int_0^{2\pi}\!\!\! \rmd \varphi \;
	\exp\mkakko{-\frac{\m_u\m_d}{2}\rme^{-i\varphi}+ \text{c.c.}}
	\!\!\!\!\!
	\\
	& \propto |\m_u \m_d|^{4N} I_0(|\m_u \m_d|) 
\ea
where $I_0$ is the modified Bessel function of the first kind. This result looks a bit pathological, because in the limit $N\to\infty$ the partition function either blows up to infinity or shrinks to zero depeding on the magnitude of $|\m_u \m_d|$. This is an inevitable consequence of the fact that only quarks of two colors participate in Cooper pairing and quarks of the third color remain gapless. They do not decouple in the low-energy limit and make it hard to rigorously define the $\epsilon$-regime for the 2SC phase. A similar situation arises in dense two-color QCD with an odd number of flavors \cite{Kogut:2000ek,Kanazawa:2009ks,Kanazawa:2009en,Akemann:2010tv}.

\section{\label{sc:conc}Summary and outlook}

In this paper, we put forward a novel RMT and analyzed its properties in the large-$N$ limit. For three flavors, we showed that the model exhibits a striking similarity to the CFL phase of QCD in that the $\SU(3)$ color is broken and gets locked to the flavor $\SU(3)$. For two flavors, we showed that an isospin-singlet pairing of quarks occurs and breaks the $\SU(3)$ color symmetry down to $\SU(2)$, in analogy to the 2SC phase of QCD. This work opens up an entirely new avenue to investigate color superconductivity in dense QCD. 

A lot of work remains to be done. As one can see from \eqref{eq:oierwsnfsdk}, in this work we have restricted ourselves to the color and flavor-antisymmetric interaction only. Extending this by incorporating the color and flavor-symmetric interaction is of much interest. It is also intriguing to numerically investigate the spectrum of the random matrix Dirac operator $\mathscr{D}$. This would be much easier (at least in the quenched limit) than analytically obtaining the microscopic spectral density, which seems to be a gargantuan task given the awful complexity of $\mathscr{D}$ in the present model.

It is well known that the CFL phase exhibits many features that match the low-density hadronic phase of QCD, and there is even a proposal that the two phases may be continuously connected \cite{Schafer:1998ef}\footnote{Recently an objection to the hadron-quark continuity was raised \cite{Cherman:2018jir}. We warn that RMT cannot be used to prove nor disprove the continuity.}. Therefore one may wonder if our new RMT can be continuously deformed into the conventional chiral RMT. Apparently this is not a trivial task, since our model requires $NM^2\sim 1$ in the large-$N$ limit whereas the conventional chiral RMT requires $NM\sim N\mu^2\sim1$, where $\mu$ is a parameter that breaks anti-Hermiticity of the Dirac operator. An insight comes from the study of RMT for dense two-color QCD \cite{Akemann:2010tv}, where it was found that results in the large-$N$ limit of \emph{strong} non-Hermiticity ($NM^2\sim\mu\sim 1$) can be recoverd from those in the limit of \emph{weak} non-Hermiticity ($NM\sim N\mu^2\sim 1$) by first taking the limit $N\to\infty$ with $N\mu^2$ fixed and then taking the second limit $N\mu^2\to\infty$. A similar methodology may work for our RMT as well. In fact, if 
\ba
	V^{A} & = - CW^{A\dagger}C
	\\
	X & = CY^\dagger C
\ea
holds, then $\mathscr{D}$ becomes anti-Hermitian. In this case we conjecture that the ordinary pattern of chiral symmetry breaking would be recovered, because the microscopic limit is universal and independent of a detailed structure of a random matrix%
\footnote{Evidence for the surprising robustness of microscopic spectral functions against deformations of the random matrix structure can be found, e.g., in Refs.~\citen{Brezin:1995dp,Jackson:1996xt,Jackson:1997ud,Vanderheyden:2002gz}.}. If this is the case, then one may softly break the above relations with a parameter $\mu$ and take the limit $N\to\infty$ with $N\mu^2\sim 1$, followed by another limit $N\mu^2\to \infty$, to connect the two regimes. This intuitive speculation must be borne out by a more detailed calculation.

\section*{ORCID}
\noindent Takuya Kanazawa - \url{https://orcid.org/0000-0002-1499-374X}

\bibliography{paper_mpla_v2.bbl}
\end{document}